\documentclass[aps,prmaterials,preprint,groupedaddress]{revtex4-2}

\usepackage{graphicx}
\usepackage{color}
\usepackage{hyperref}\hypersetup{colorlinks=true,linkcolor=blue,citecolor=blue}
\usepackage{siunitx}
\usepackage{stmaryrd}	
\usepackage{multirow}
\usepackage{bm}
\usepackage{amssymb,amsmath,amsfonts,latexsym,mathrsfs,amsthm}
\usepackage{soul}

\begin{document}

\title{Electric leakage suppression of phase-transforming ferroelectrics with donor impurities}

\author{Chenbo Zhang}
\email{cbzhang@tongji.edu.cn}
\affiliation{\small MOE Key Laboratory of Advanced Micro-Structured Materials, School of Physics Science and Engineering, Institute for Advanced Study, Tongji University, Shanghai 200092, China}

\author{Xiaotong Peng}
\affiliation{\small MOE Key Laboratory of Advanced Micro-Structured Materials, School of Physics Science and Engineering, Institute for Advanced Study, Tongji University, Shanghai 200092, China}

\author{Bo Liu}
\affiliation{\small MOE Key Laboratory of Advanced Micro-Structured Materials, School of Physics Science and Engineering, Institute for Advanced Study, Tongji University, Shanghai 200092, China}

\author{Kai Zhang}
\affiliation{\small  School of Aerospace Engineering and Applied Mechanics, Tongji University, 200092 Shanghai, China}

\author{Xian Chen}
\email{xianchen@ust.hk}
\affiliation{\small Department of Mechanical and Aerospace Engineering, Hong Kong University of Science and Technology, Clear Water Bay, Hong Kong}

\date{\today}

\begin{abstract}
\noindent 

Phase-transforming ferroelectric materials are widely used in energy harvesting and conversion devices. However, the functionality of these devices is significantly impeded by electrical leakage at high temperatures. In this study, we fundamentally study the mechanism of electrical leakage suppression due to phase transformation in a series of donor-doped ferroelectric oxides, Ba$_{0.955}$Eu$_{0.03}$Ti$_{1-x}$Zr$_{x}$O$_{3}$ with $0\leq x\leq 0.15$. Our experiments clearly demonstrate that the symmetry-breaking phase transformations result in the reduction in electrical conductivity of the donor-doped ferroelectric oxides. The DFT calculation suggests that the donor energy level undergoes a shallow-to-deep transition at the phase transformation temperature. By analyzing the constitutive model of the leakage current density function, we propose a leakage suppression coefficient that rationalizes the development of ferroelectrics with low electrical leakage at elevated temperatures.

\end{abstract}

\maketitle

\section{Introduction}
For multiferroic materials, the transport properties such as polarization, magnetization, heat capacity and strain are very sensitive to the symmetry-breaking phase transformation. Changes in lattice parameters usually result in an abrupt jump of one or more ferroic properties. This phenomenon paves a way for the new types of energy conversion applications such as pyroelectric generators and elasto/magneto/electrocaloric cooling devices {\cite{bucsek2019direct,zhang2019power,li2020effect,fahler2018caloric,hou2018ultra}. Regarding energy applications of these dielectric materials, the increasing electric leakage at elevated temperature limits the device performance and functionalities. Conventionally, the electric resistance and temperature are inversely related. In some ferroic materials such as barium titanates and potassium niobates, a sudden decrease in electrical conductivity near the phase transition temperature has been observed, indicating the suppression of electrical leakage at high temperatures \cite{lewis1985ptcr, zhu2014ptcr}. This unusual temperature-dependent conductivity in these materials was understood as the effect of grain boundaries in many relevant works \cite{bell2021barium,heywang1964resistivity,goodman1963electrical,prohinig2021modification,bell2021barium,holsgrove2017mapping}.

 A recent study shows that by a systematic lattice parameter tuning for barium titanates with similar grain morphology and equivalent grain boundaries, the leakage current density can be suppressed by an order of magnitude in the regime of solid-solid phase transformation \cite{zhang2020impact}. 
 This suggests a fundamental leakage suppression mechanism, which should be associated with the change of crystal structures during the phase transformation. Table \ref{tab_PTCR} presents similar examples. These materials demonstrate a rise in electrical resistivity following symmetry-breaking phase transformations. However, the mechanism underlying this coupling between structure and property is not yet fully understood. This paper aims to fundamentally explore the connection between the electrical structure of ferroelectrics and the alterations in crystal structures during phase transformation.

\begin{table*}[ht]
	\centering
	\caption{Ferroelectric single crystals with leakage current suppression at the transformation temperature.\label{tab_PTCR}}
	{\small
		\begin{ruledtabular}
			\begin{tabular}{>{\centering\arraybackslash}p{ 0.35\textwidth}|>{\centering\arraybackslash}p{0.24\textwidth}|>{\centering\arraybackslash}p{0.15\textwidth}|>{\centering\arraybackslash}p{0.15\textwidth}|c}
				\multirow{2}{*}{Single crystals}    & Trans.  & \multicolumn{3}{c}{Electrical resistivity}\\ \cline{3-5}
				& temperature & before & after & unit  \\ \hline
				
				BaTiO$_3$\cite{kawabe1963resistivity} & 120$^\circ$C & 10$^{8}$ & 10$^{10}$ & $\Omega$m\\
				(Nb,Ba)TiO$_3$\cite{brown1964electrical} & 122$^\circ$C & $\sim$0.5& 10 & $\Omega$m\\
				(La,Ba)TiO$_3$\cite{motohira1996single} & 110$^\circ$C & 0.7& 1.1 & $\Omega$m\\
				(Fe,Ba)TiO$_3$\cite{yamato2007universality} & 120$^\circ$C & 0.5 & 3 & $n\Omega$cm\\
				(Pb(Zn$_{1/3}$Nb$_{2/3}$)O$_{3}$)$_{0.92}$ & \multirow{2}{*}{160$^\circ$C }& \multirow{2}{*}{50} & \multirow{2}{*}{100} & \multirow{2}{*}{$n\Omega$cm}\\
				-(PbTiO$_{3}$)$_{0.08}$\cite{yamato2007universality} && & & \\
			\end{tabular}
		\end{ruledtabular}
	}
\end{table*}

The relationship between multiferroic properties and electronic structures has been widely studied in experiments. It has been demonstrated that the electrical conductivity in ferroelectric materials can be modulated by adjusting the energy band gap between the conduction and valence bands \cite{yang2019improved,sheeraz2019enhanced,prajapati2022band,shi2019enhanced}. The unusual suppression of leakage in ferroelectric materials was observed with the presence of donor impurities \cite{chen2011ptcr,wang2018influence,pithan2020defect}. These impurities contribute to the formation of n-type semiconducting ferroelectrics, the energy levels of which are highly sensitive to the lattice parameters of the crystal. Most barium titanate-based ferroelectrics undergo a first-order phase transformation, leading to sudden changes in lattice parameters. It is crucial to investigate how these changes in lattice parameters affect the electronic structures, thereby unveiling the mechanism behind the abnormal suppression of leakage during phase transformation in ferroelectric oxides.

\section{Material development}

We design a donor-doped barium titanate ferroelectric oxide with 3 at\% A site doping by Eu$^{3+}$ cation. This design approximately creates a charge vacancy per 33 unit cells. The B site dopant, Zr$^{4+}$ with varying compositions is added to finely tune the lattice parameters of both ferroelectric and paraelectric phases \cite{zhang2021energy}. A series of donor-doped ferroelectric oxides can be presented as Ba$_{0.955}$Eu$_{0.03}$Ti$_{1-x}$Zr$_{x}$O$_{3}$ (i.e. Eu-BTO-Zr$_x$), in which the atomic percentages of Ba$^{2+}$ and Eu$^{3+}$ are calculated based on the charge balance. A family of Eu-BTO-Zr$_x$ were synthesized systematically for $0\leq x \leq 0.15$. The grain morphology is engineered to be comparable for all compositions, so that we can focus on the influence of fundamental mechanism due to the donor impurity. We also utilize the first-principle calculations to investigate the band structures of the donor-doped barium titanates, and study the influences of variation of lattice parameters on electric conductivity during the phase transformation.

\begin{figure*}[ht]
	\includegraphics[width = 1\textwidth]{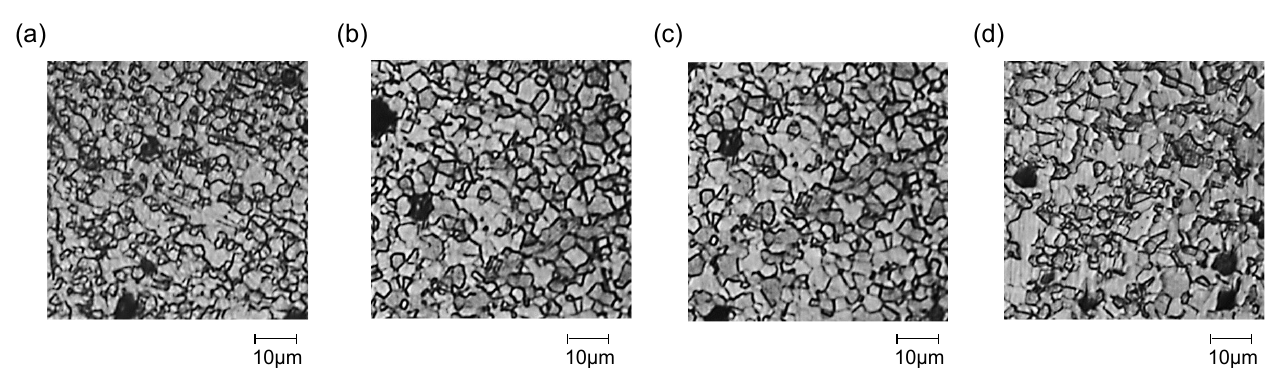}
	\caption{ Grain morphology analysis of ferroelectric Eu-BTO-Zr$_x$ system. Optical micrographs of (a)
		Zr$_{0}$, (b) Zr$_{0.05}$, (c) Zr$_{0.10}$, (d) Zr$_{0.15}$ sample after chemical etching.
	}\label{Fig_grain}
\end{figure*}

The Eu-BTO-Zr$_x$ material is synthesized by the solid-state reaction method. The same temperature and duration condition  were applied to assist the sintering process so that a comparable grain morphology for all developed Zr compositions can be achieved. The synthesis procedures follow the recent works on grain-size effect of Sr doped barium titanates \cite{zhang2021low}. After sintering, the donor-doped ferroelectric specimen was polished and etched to reveal the grain morphology, as shown in Fig. \ref{Fig_grain}. 
The microstructure was measured by the differential interference microscope \cite{zeng2019quantitative}. The phase contrast among various grains is associated with the crystal orientation. From the morphological features illustrated in Fig. \ref{Fig_grain}, the average grain size is approximately 5$\mu$m with a uniform orientation distribution for all Zr compositions from 0 to 0.15. 

\section{Electric leakage suppression at phase transformation} \label{sec:char}

The electric transport properties of Eu-BTO-Zr$_x$ specimen were characterized by aix ACCT TF2000E ferroelectric analyzer, while the thermal properties of phase transformation were measured by differential scanning calorimetry (i.e. Thermal Analysis Instrument DSC-250) from 40$^\circ$C to 120$^\circ$C. Fig.~\ref{Fig_leak}(a)-(b) shows the temperature dependent leakage current density and dielectric constant for Zr composition varying from 0 to 0.15. From DSC measurements in Fig. \ref{Fig_leak}(c), the first-order phase transformation occurs in the specimen with Zr$_0$ and Zr$_{0.05}$. The other two compositions Zr$_{0.10}$ and Zr$_{0.15}$ do not exhibit any signs of phase transformation in this temperature range. In the vicinity of the transformation temperature, the leakage current density started tapering off (i.e. gradually reducing) the monotonic increasing trend. Especially for the Zr$_{0.05}$ composition, the leakage current density plateaued around 70$^\circ$C.

\begin{figure}[ht]
	\includegraphics[width = 0.5\textwidth]{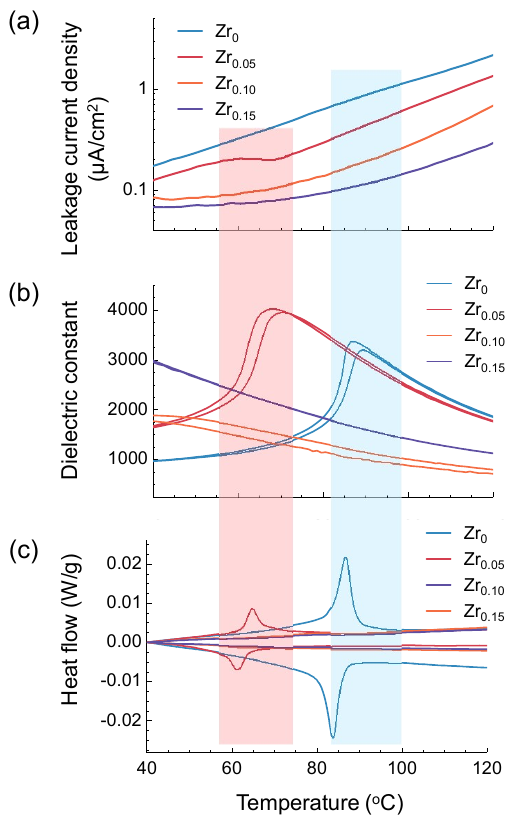}
	\caption{Temperature dependent (a) leakage current density, (b) dielectric constant and (c) heat flow in ferroelectric Eu-BTO-Zr$_x$ system.}\label{Fig_leak}
\end{figure}

The transport property, i.e. dielectric constant, experiences an abrupt increase at the phase transformation temperatures for Zr$_{0}$ and Zr$_{0.05}$ respectively, in Fig.\ref{Fig_leak}(b). Compared to the phase-transforming ferroelectrics, the nearby compositions of Zr$_{0.10}$ and Zr$_{0.15}$ do not show any anomalous trend of the transport properties even with the same atomic percentage doping by Eu$^{3+}$ donor. Our experiments suggest that the formation of donor point defect is not sufficient to trigger the leakage suppression in ferroelectric oxides. The change of crystal structures associated with first-order phase transformation is coupled with the effect of donor point defect to influence the electric conductivity. 

\begin{figure}[ht]
	\includegraphics[width = 0.5\textwidth]{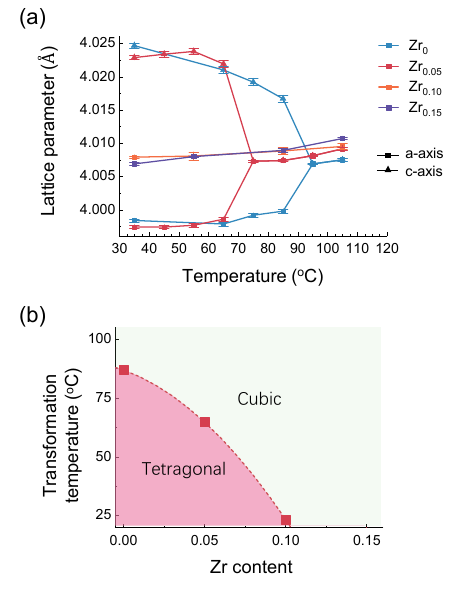}
	\caption{(a)Temperature dependent lattice parameters and (b) composition driven phase transition in ferroelectric Eu-BTO-Zr$_x$ system.}\label{Fig_xrd}
\end{figure}

We characterized the crystal structure and lattice parameters of Eu-BTO-Zr$_x$ specimen by X-ray powder diffraction (Panalytical X’pert pro diffractometer) from 30$^\circ$C to 110$^\circ$C. The changes of lattice parameters are presented in Fig. \ref{Fig_xrd}(a), which were determined and refined by Rietveld method \cite{rodriguez1993recent}. For Zr$_{0}$ and Zr$_{0.05}$ specimen, the structural transformation from tetragonal phase to cubic phase are clearly characterized as the sudden collapse of tetragonality near the transformation temperatures at 70$^\circ$C and 110$^\circ$C respectively, while the lattice parameters of the other two compositions simply followed the trend of thermal expansion with the cubic symmetry throughout the tested temperature range. A phase diagram is constructed in Fig. \ref{Fig_xrd}(b) by integrating the crystal structural parameters and corresponding transformation temperatures of the Eu-BTO-Zr$_x$ family. With increasing Zr compositions in the donor-doped barium titanates, the phase diagram suggests that the allowable temperature range for thermally driven phase transformation gets narrower, by which the development of the donor-doped barium titanates in the compositional neighborhood of Zr$_{0.05}$ is possible to further enhance such a suppression of leakage.

\section{Transition of defect energy level}

From the experimental perspectives, the suppression of electric conductivity is related to the change of lattice parameters in the donor-doped barium titanates. At phase transformation, the abrupt change of lattice parameters influence the transport properties coupled with the effect on energy levels through donor point defects in the crystal. We perform the first-principle calculation to investigate the change of band structure during phase transformation for donor doped barium titanates. The calculation is conducted based on the density functional theory (DFT) with a projector augmented wave method \cite{kresse1999ultrasoft} as
implemented in the Vienna Ab initio Simulation Package (VASP) software \cite{kresse1993ab,kresse1996efficiency,kresse1996software}. 
The exchange-correlation functional is described by Perdew-Burke-Ernzerhof (PBE) \cite{perdew1996generalized}. For the doped atomic model, we adapt the virtual crystal approximation method \cite{bellaiche2000virtual} to represent the partial occupation of 0.05, 0.10, 0.15 Zr doping, which achieves the same stoichiometric ratio as in experiments. The Eu composition is achieved by atomic substitution in the designed supercell.  The automatically generated uniform Monkhorst-Pack grid is used to achieve the 0.15 $\AA^{-1}$ K-point distance, which ensures sufficient calculation accuracy. The convergence criterion is 10$^{-6}$. To study the band structures at different temperatures, we utilize the experimentally characterized lattice parameters in the DFT simulation to obtain the influence of variable structural parameters on the band structures. Similar approach is introduced in reference \cite{vzelezny2015variation}. The atomic position is relaxed to achieve the stable structure, used for band structure calculation. The spin polarization and additional Hubbard-U terms are not considered to save computational cost, where the calculated band gap is notably lower than the experimental value. 

\begin{figure}[ht]
	\includegraphics[width = 0.7\textwidth]{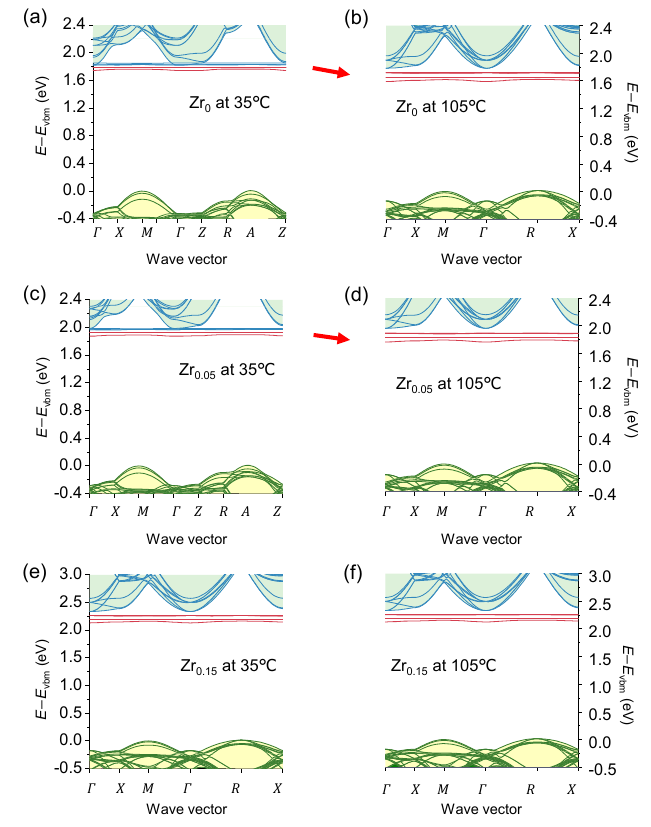}
	\caption{Band structure in ferroelectric Eu-BTO-Zr$_x$ system at various temperatures: (a) Zr$_{0}$, (c) Zr$_{0.05}$, (e) Zr$_{0.15}$ at 35$^\circ$C; (b)Zr$_{0}$ ,(d)Zr$_{0.05}$, (f)Zr$_{0.15}$ at 105$^\circ$C.}\label{Fig_band}
\end{figure}

Fig.~\ref{Fig_band}(a)-(f) shows the band structures of Zr$_{0}$, Zr$_{0.05}$ and Zr$_{0.15}$ along high symmetry points at various temperatures. The shaded blue and yellow energy levels are conduction and valence band respectively. In the energy gap between conduction and valence band, all compositions present defect energy levels, which is close to the edge of conduction band. By comparison of band structure at low and high temperature, the defect energy levels in Zr$_{0}$ and Zr$_{0.05}$ at high temperature (cubic phase) shifts away from the conduction band, which becomes deeper than low temperature (tetragonal) phase. Such a shallow-to-deep donor transition does not appear in Zr$_{0.15}$ composition. Those results suggest the significant influence of phase transformation on defect energy levels.

In dielectric materials, the electric conductivity, i.e. leakage current density, strongly replies on the energy gap (E$_g$) between conduction band and valence band. We calculated the band gap E$_g$ for phase-transforming Zr$_0$, Zr$_{0.05}$ and non-transforming Zr$_{0.15}$ at varying temperatures from 30$^\circ$C to 110$^\circ$C, as seen in Fig. \ref{Fig_dos}(a).  As Zr composition increases, the band gap gets larger, which agree well with the trend of leakage current density measured experimentally in Fig.~\ref{Fig_leak}(a). 

\begin{figure*}[ht]
	\includegraphics[width = 1\textwidth]{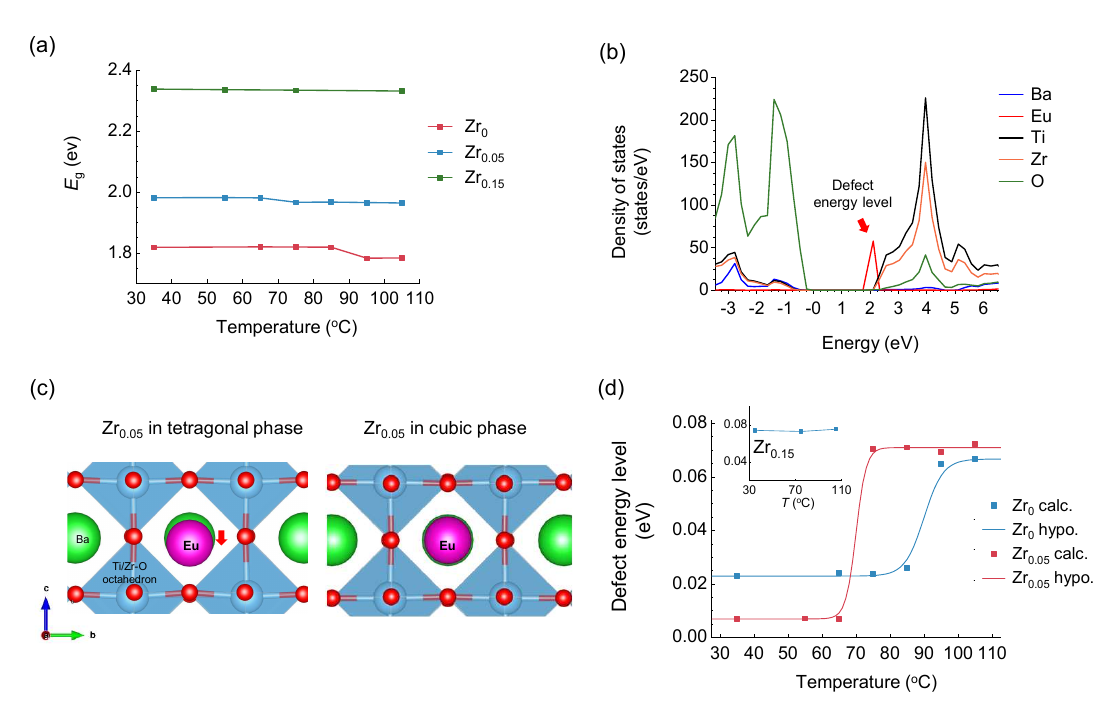}
	\caption{(a) Temperature dependent band gap; (b) projected density of states; (c) relaxed atomic position at different crystal structure; (d) temperature dependent defect energy levels in ferroelectric Eu-BTO-Zr$_x$ materials.  }\label{Fig_dos}
\end{figure*}

For phase-transforming Zr$_0$ and Zr$_{0.05}$, both exhibit a small drop of band gap energy at their transformation temperatures respectively. The experimental results in Section \ref{sec:char} evidently corresponds this phenomenon to a gradual reduction of conductivity. We conjecture that the defect energy level plays a role to suppress the electric leakage during the phase transformation. In Fig. \ref{Fig_dos}(b), we studied the defect energy levels through the analysis of the orbital projected density-of-states in Zr$_{0.05}$ at 35$^\circ$C. It is seen that the defect energy levels generated by Eu$^{3+}$ donor are localized within the band gap. As a consequence of the donor, the energy levels are hybridized by Eu-$f$, Ti-$d$, Zr-$d$ and O-$p$ orbitals. In Fig.~\ref{Fig_dos}(c), the relaxed atomic position of Eu shifts at different crystal structure. Across phase transformation, the evolution of lattice parameter changes the local bonding environment of Eu, which tunes the defect energy level.  We extract the defect energy levels at different temperatures in Zr$_{0}$, Zr$_{0.05}$ and Zr$_{0.15}$ compositions, as shown in Fig.~\ref{Fig_dos}(d). Both Zr$_{0}$ and Zr$_{0.05}$ compositions exhibits a jump of defect energy level across their phase transformations. The defect transforms into deeper states after transformation from tetragonal to cubic phase. In the Zr$_{0.15}$ composition, there is no phase transformation and the defect energy level at different temperatures are same. The results tells the leakage suppression at phase transformation is closely related to the change of defect energy levels. 

\section{Further analysis}
\subsection{Constitutive relation}
Doping heterovalent ions and dopants with different ionic radius generate large lattice distortion, in which both cations and anions deviate from regular lattice sites. We analyze the temperature dependent leakage current density $J(E,T)$ referring to the Poole-Frenkel controlled leakage current \cite{mccormick2003microstructure},
\begin{equation}
  J(E,T)=\mu E \text{e}^{(-\frac{E_a}{k_BT}+\chi E^{\frac{1}{2}}})\label{eqn_arrhe},
\end{equation}
in which $E$ is the applied electric field in leakage characterization, $k_{B}$ is the Boltzmann constant, $T$ is the temperature in Kelvin, $E_a$ is the activation energy. The $\mu$ and $\chi$ are constants. Based on the electronic band structure in Fig. \ref{Fig_band}, the heterovalent Eu dopant forms the shallow donor state following the electron-impact ionization mechanism of cation impurity \cite{williams2006threshold}, in which the activation energy is proportional to the ionization energy of defect energy level. The activation energy $E_a$ is defined as
\begin{subequations}
\begin{align}
        E_a&=E_a^0+\zeta \epsilon(T),\label{eqn_Ea}\\
        \zeta&= \dfrac{\rho_d}{\rho_{c}},\label{eqn_zeta}
\end{align}
\end{subequations}
where $E_a^0$ is the initial activation energy, $\epsilon(T)$ is the defect energy level across phase transformation, $\zeta$ is a proportional constant that describes the ratio of occupied states by the defect ($\rho_d$) to the conduction band ($\rho_{c}$), which is calculated by their density of states. Based on the abrupt change of defect energy level in Fig.~\ref{Fig_dos}(d), the $\epsilon(T)$ across phase transformation is proposed as
\begin{equation}
  \epsilon(T)= \epsilon_p+\dfrac{\epsilon_f-\epsilon_p}{1+\text{e}^{\frac{T-T_c}{\kappa}}},\label{eqn_sigmoid}
\end{equation}
in which $\epsilon_f$ and $\epsilon_p$ is the defect energy level at ferroelectric phase and paraelectric phase, $T_c$ is the transformation temperature, $\kappa$ represents the derivative of the temperature dependent defect energy level. The hypothetical model is plotted as the dashed lines in Fig.~\ref{Fig_dos}(d), which well capture the feature of change of defect energy level across the phase transformation for donor-doped ferroelectrics. 

\squeezetable
\begin{table}[ht]
	\caption{Parameters used in hypothesized constitutive model of leakage current density for Zr$_{0.05}$ composition.\label{tab_para}}
	{\small
		\begin{ruledtabular}
			\begin{tabular}{ccc}
				Variable & Value & Unit\\
				\hline
				$\mu$ & 4.71    &$\mu$A/cm$^2$V \\
				$\epsilon_f$ & 0.007 & eV\\
				$\epsilon_p$ & 0.072 & eV\\
				$T_c$ & 338 & K\\
				$\zeta$ & 0.25 &- \\
				$E_a^0$ &0.36&eV\\
				$\chi$ & 0.003& cm$^{1/2}$/kV$^{1/2}$\\
				$\kappa$ & 1.50 & eV/K
			\end{tabular}
		\end{ruledtabular}
	}
\end{table}

By equation (\ref{eqn_arrhe})-(\ref{eqn_sigmoid}), we calculate the leakage current density of Zr$_{0.05}$ under different temperatures and electric fields. The parameters in the constitutive model are listed in Table \ref{tab_para}. Fig.~\ref{Fig_const}(a) shows that the hypothetical model well predicts the trend of leakage suppression at phase transformation. The amplitude of suppressed leakage agrees well with the experiments.

\begin{figure*}[!]
	\includegraphics[width = 1\textwidth]{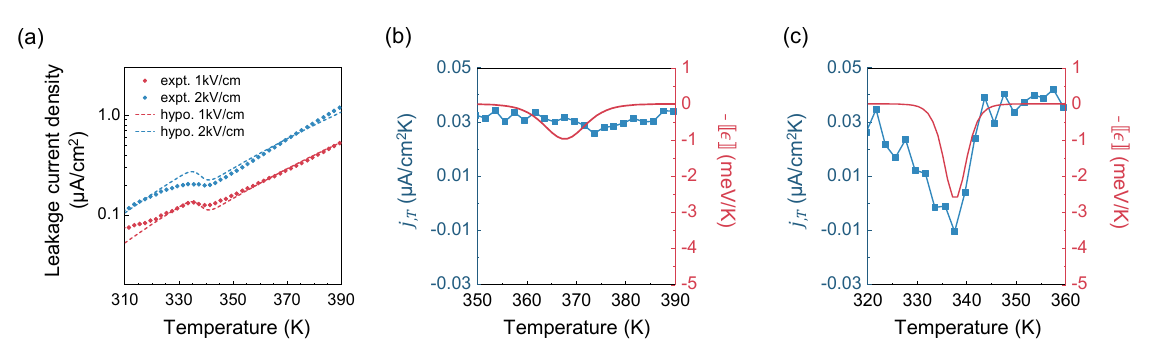}
	\caption{(a) Constitutive model of leakage current density in Zr$_{0.05}$ composition; correlation between the coefficient of leakage suppression at phase transformation and temperature partial derivative of leakage current density in : (b) Zr$_0$ and (c) Zr$_{0.05}$ composition in ferroelectric Eu-BTO-Zr$_x$ materials.  }\label{Fig_const}
\end{figure*}

\subsection{Coefficient of leakage suppression at phase transformation}

To reveal the dominant temperature dependent factor on leakage suppression at phase transformation, we express the nature logarithm of \eqref{eqn_arrhe} as, 
\begin{equation}
	j(E,T)=\mu E \left(-\frac{E_a^0+\zeta \epsilon(T)}{k_BT}+\chi E^\frac{1}{2}\right), \label{eqn_lnFrenkle}
\end{equation}
where $j(E,T)=\text{ln}J(E,T)$. The constitutive model shows the leakage current density depends on both electric field and temperature. We analyze the effect of thermally driven phase transformation and the temperature partial derivative of $j$, giving
\begin{equation}
	j_{,T}=\mu E \left[ \frac{E_a^0}{k_B}T^{-2}+\frac{\zeta}{k_B}\epsilon(T)T^{-2}-\frac{\zeta}{k_B}\epsilon_{,T}T^{-1}\right]. \label{eqn_deriv}
\end{equation}
During the phase transformation, the partial derivative $j_{,T}$ can be expanded via Taylor series as
\begin{equation}
    j_{,T} = j_{,T}(T_c)+j_{,TT}(T_c)(T-T_c)+o(T-T_c), \label{eqn_taylor}
\end{equation}
in which $o(T-T_c)$ is the high order term. The first derivative $j_{,T}(T_c)$ and second derivative $j_{,T}(T_c)$ are obtained from equation (\ref{eqn_deriv}). We ignore the high order term and substitute the $\epsilon(T)$ from equation (\ref{eqn_sigmoid}), the equation (\ref{eqn_taylor}) is re-written as
\begin{equation}
\begin{split}
        j_{,T} = [E_a^0+\zeta \epsilon(T_c)]\alpha(T)+
        \epsilon_{,T}(T_c)\beta(T)+\epsilon_{,TT}(T_c)\gamma(T)\label{eqn_reformderiv}
\end{split}
\end{equation}
in which 

\begin{subequations}
\begin{align}
        \alpha(T) &= \frac{\mu E}{k_{B}}(3T_c^{-2}-2T_c^{-3}T),\\
        \beta(T) &=\frac{\mu E\zeta}{k_{B}}(-3T_c^{-1}+2T_c^{-2}T),\\
        \gamma(T) &=\frac{\mu E\zeta}{k_{B}}(-T_c^{-1}T+1).
\end{align}
\end{subequations}
At transformation temperature, $T\rightarrow T_c$. For most barium titanates based ferroelectrics, $T_c > 300$K, thus $1/T_c^2 \approx 0$. An estimate of the coefficients at phase transformation are
\begin{subequations}
\begin{align}
        \alpha(T) & \approx  0,\\
        \beta(T) & \approx -\frac{\mu E\zeta}{k_{B} T_c},\\
        \gamma(T) & \to 0.
\end{align} \label{eqn_simpl}
\end{subequations}
By equation (\ref{eqn_reformderiv}) and (\ref{eqn_simpl}), we have
\begin{equation}\label{eq:jt}
   j_{,T}(T_c) = -\zeta\epsilon_{,T}(T_c)\frac{\mu E}{k_{B}T_c}. 
\end{equation}
The function \eqref{eq:jt} determines the leakage suppression mechanism during phase transformation for ferroelectric materials with donor point defects. That is, the amplitude of leakage suppression at transformation temperature is proportional to the derivative of defect energy level with respect to temperature. By equation (\ref{eqn_zeta}), we propose a coefficient of leakage suppression at phase transformation as 
\begin{equation}
	\llbracket \epsilon \rrbracket = \frac{\rho_d}{\rho_c} \epsilon_{,T}(T_c).
\end{equation}
We calculate this coefficient in Zr$_{0}$ and Zr$_{0.05}$, which is shown in Fig. \ref{Fig_const}(b) and (c). At phase transformation temperature, the coefficient of leakage suppression is 2.6 meV/K in Zr$_{0.05}$, which is 3 times higher than Zr$_{0}$.

\section{Conclusion}
In summary, we explore the mechanism of suppressed electric leakage coinciding with the symmetry-breaking phase transformation in a family of Eu-BTO-Zr$_{x}$ ferroelectric materials with comparable grain morphology. Our findings reveal that the electric leakage can be manipulated by tuning lattice parameter at varying Zr compositions, in which the Zr$_{0.05}$ composition exhibits significant suppression of leakage at phase transformation. We discover that the Eu donor energy level undergoes a shallow-to-deep transition across the phase transformation. We theorize the impact of defect energy level on temperature dependent leakage current density and propose a coefficient of leakage suppression at phase transformation. As a result, this coefficient in Zr$_{0.05}$ is three times larger than Zr$_{0}$ composition. Our study reveals the fundamental role of the impact of phase transformation on leakage suppression of ferroelectric materials with donor hybridization. Our work underlies a rational approach to develop high-performance phase-transforming ferroelectric materials for energy harvesting and conversion devices.

\section{Acknowledgments}
C. Z. acknowledge the support by National Natural Science Foundation of China (No.12204350), Fundamental Research Funds for the Central Universities (No.22120240009). X. C. thank the financial support under GRF Grants 16203021, 16204022, 16203023 and CRF Grant No. C6016-20G-C by Research Grants Council, Hong Kong. X.C. and C.Z. thank the support by Innovation and Technology Commission, HK through the ITF Seed grant ITS/354/21.

\bibliographystyle{apsrev4-2}
\bibliography{Citation.bib}

\end{document}